\documentclass[runningheads]{llncs}
\usepackage{graphicx}
\usepackage{booktabs}
\usepackage[table,xcdraw]{xcolor}
\usepackage{multirow}
\usepackage{tabularx}
\usepackage{hyperref}
\usepackage{parskip}

\begin{document}
\title{Prototype of a Cardiac MRI Simulator for the Training of Supervised Neural Networks}
\titlerunning{Cardiac MRI Simulator}
%
\author{Marta Varela\inst{1}\orcidID{0000-0003-4057-7851} \and
Anil A Bharath\inst{2}\orcidID{0000-0001-8808-2714}}
\authorrunning{M. Varela et al.}
%
\institute{National Heart \& Lung Institute, Imperial College London, London, UK
\email{marta.varela@imperial.ac.uk}\\
\and
Bioengineering Department, Imperial College London, London, UK\\
\email{a.bharath@imperial.ac.uk}}
\maketitle              
\begin{abstract}
Supervised deep learning methods typically rely on large datasets for training. Ethical and practical considerations usually make it difficult to access large amounts of healthcare data, such as medical images, with known task-specific ground truth. This hampers the development of adequate, unbiased and robust deep learning methods for clinical tasks.

Magnetic Resonance Images (MRI) are the result of several complex physical and engineering processes and the generation of synthetic MR images provides a formidable challenge. Here, we present the first results of ongoing work to create a generator for large synthetic cardiac MR image datasets. As an application for the simulator, we show how the synthetic images can be used to help train a supervised neural network that estimates the volume of the left ventricular myocardium directly from cardiac MR images.

Despite its current limitations, our generator may in the future help address the current shortage of labelled cardiac MRI needed for the development of supervised deep learning tools. It is likely to also find applications in the development of image reconstruction methods and tools to improve robustness, verification and interpretability of deep networks in this setting.

\end{abstract}

\section{Introduction}
Recent developments in machine learning (ML) have improved rapid reconstruction techniques, enabled automatic image analysis and enhanced the interpretation of medical images in general and cardiac Magnetic Resonance Imaging (MRI) in particular \cite{Litjens2019State-of-the-ArtAnalysis}. Progress in this area has nevertheless been restricted by shortages in large anonymised curated datasets, and difficulties in obtaining high-quality ground truth annotations for supervised learning tasks\cite{VaroquauxMachineFuture}. These problems are compounded by the under-representation of patients from minority backgrounds and those with infrequent anatomical variations or rare diseases. Moreover, deploying ML models to novel imaging sequences is often delayed until a large numbers of similarly parameterised images have been accrued.

The creation of large datasets of synthetic images whose properties follow prescribed statistical distributions could help alleviate some of the current constraints in the deployment of supervised neural networks (NNs) for medical imaging tasks. The physical processes underlying the nuclear magnetic resonance (NMR) of water protons in biological tissue are complex, as are the interactions of the protons' magnetization vectors with the MRI magnets and other engineering equipment. As such, simulating MRI acquisition necessarily involves simplifications and trade-offs between accuracy and speed. Several simulators with varying degrees of complexity and focusing on different anatomical regions have been proposed \cite{Benoit-Cattin2005TheSimulator,Stoecker2010High-performanceSimulations,Kose2017BlochSolver:Sequences,Xanthis2019CoreMRI:Cloud,Bittoun1984AMethod}. So far none of these has allowed the automatic generation of large sets of MR images with controlled variation in anatomical or imaging parameters suitable for the training of NNs.

\subsubsection{Aims}
We present a generator of synthetic cardiac MR images, designed to allow the creation of large imaging datasets with controlled parameter variations. As an application, we show how the synthetic data can be used to train a network that estimates left ventricular myocardial volume (LVMV).

\section{Methods}
\subsection{MRI Simulator}
We created a Python 3.8-based modular simulator of cardiac MRI. The simulator takes the following information as independent inputs:
\begin{enumerate}
    \item Cardiac phantom (described below);
    \item Scanner characteristics. We assumed simulated image acquisition at a $B_0=1.5$ T, with a spatially uniform $B_1$ field and perfect gradient coils with an infinite slew rate. We also assumed a single receive coil with a spatially uniform sensitivity.
    \item List of MR sequence parameters similar the ones inputted by radiographers at the time of scanning. We used an axial 3D gradient echo sequence, with: an echo time of 40-60ms, a repetition time of 3000ms, an excitation flip angle of $10-20^\circ$ and a 80 MHz bandwidth. We used a Cartesian sampling scheme with a linear phase encode order, RF spoiling and no parallel imaging capabilities. 
\end{enumerate}
We solved the Bloch equations with the CPU-based forward Euler single-compartment solver initially proposed by Liu \textit{et al.} \cite{Liu2017FastModel} and a fast Fourier transform for image reconstruction. Spatial motion of the excited protons during the image acquisition process (caused by blood flow, cardiac and respiratory cycles, diffusion or patient motion) was not simulated.

\subsection{Cardiac MRI Phantom}
Digital cardiac phantoms are a necessary component of cardiac medical image simulators. There are several atlases and mesh-based models of the human heart \cite{Young2009}, but realistic cardiac MR images rely on accurate representations of the entire thoracic anatomy. Digital models that rely on non-uniform rational B-splines (NURBS) are particularly flexible and efficient and have been employed in computed tomography and nuclear medicine image simulators in the past. \textbf{blue}{The different approaches employed to generate computational human phantoms are carefully reviewed elsewhere \cite{Kainz2019}}.

We used the XCAT phantom \cite{Segars2010,Segars2018ApplicationBeyond}, which is a detailed NURBS-based representation of human anatomy originally based on the segmentations of the high-resolution Visible Human Male and Female images. We took the thoracic region of the XCAT phantom adjusted to $50^{th}$-percentile organ volume values \cite{Segars2018ApplicationBeyond} as our baseline anatomical representation. Taking advantage of NURBS's flexibility in representing structures with varying morphology and volume, we independently varied the following anatomical parameters: heart dimensions (varied independently in the FH, LR and AP directions); LV radius; apico-basal length; LV thickness; LV thickness close to the apex. All these parameters followed independent uniform distributions, so that their value varied between 80\% and 120\% of the original value in the reference XCAT phantom. For each instance, we randomly chose between the male or female phantom representation. Clinical experts visually inspected the generated phantom anatomy instances to ensure that they were anatomically plausible (i.e., that they could be segmentations of realistic torso anatomies). We also confirmed that the LVMV range in the generated phantoms was similar to the LVMV in the image patients (see below). This was the only quantitative test of anatomical plausibility we performed on the simulated images. 

\begin{figure}[!h]
\centering
\includegraphics[width=1.0\textwidth]{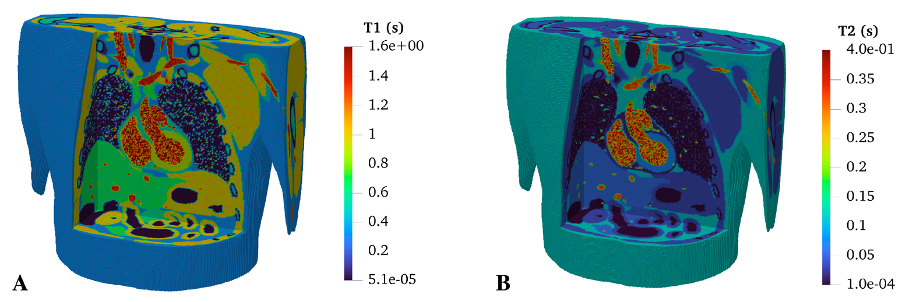}
\caption{Cross section of the $T_1$ (A) and $T_2$ (B) maps for the XCAT phantom. The values assigned to each voxel follow a normal distribution around the literature values of the respective
tissue. The standard deviation is 30\% for blood and 10\% for all other tissues. Lumen and cortical bones were treated like air and assigned proton densities close to zero. A similar approach was followed when assigning $T_2^*$ values.}
\label{fig:XCAT}
\end{figure}

We then voxelised the phantom representations into images with dimensions $256 \times 256 \times 15$ with a $1.0 \times 1.0 \times 6.0~mm^3$-resolution. Before voxelisation, we randomly introduced translations of $<2~cm$ and rotations of $<6^\circ$ along any axis to the phantom in order to model the variability of body position within the scanner. For each of the 17 segmented tissues in the phantom (see Supplementary Table 1), we created Gaussian distributions of water proton NMR relaxation time constants ($T_1$, $T_2$, $T_2^*$). The distributions were centred on literature values at $1.5~T$ for each NMR parameter and their standard deviation was set to 10\% of the corresponding literature value. An exception was made for arterial and venous blood, whose standard deviation was set to 30\% to simplistically model the greater variability in signal due to blood flow. We neglected variations in magnetic susceptibility across the body. Each voxel in the simulated image had NMR parameters randomly chosen from the multivariate Gaussian distribution corresponding to its tissue type. 


\subsection{Left Ventricular Myocardial Volume (LVMV) Estimation}
To demonstrate the potential application of the cardiac MRI simulator as a generative model for deep learning applications, we create an \textit{in silico} dataset of 500 $T_2$-weighted axial stacks of cardiac images, with variations in phantom anatomy and MR sequence parameters detailed above.
We train a regression NN to automatically estimate the volume of the LV myocardium (VLVM) in three different experimental setups - see Table \ref{tab:experiments}. 
The patient data was part of an ethically approved retrospective study. It was acquired in a 1.5T Siemens scanner, using a $T_2$-prepared multi-slice gradient echo sequence (TE/TR: 1.4/357 ms, FA: $80^\circ$, resolution: $1.4 \times 1.4 \times 6~mm^3$). Ground truth VLVMs are calculated using an existing segmentation CNN \cite{Howard2021a} applied to the patient images. 

\begin{table}[]
\centering
\begin{tabular}{|c|c|c|}
\hline
\rowcolor[HTML]{FFEBCD} 
\textbf{Experiment}                          & \textbf{\# Training Images} & \textbf{\# Test Images} \\ \hline
\cellcolor[HTML]{FFEBCD}\textbf{A} &        400 (simulated)      &    100 (simulated)      \\ \hline
\cellcolor[HTML]{FFEBCD}\textbf{B} &        393 (patient)        &    99 (patient)         \\ \hline
\cellcolor[HTML]{FFEBCD}\textbf{C} &        393 (patient) + 500 (simulated) &  99 (patient) \\ \hline
\end{tabular}%
\caption{Number of simulated and patient images used in each of the experiments, as part of the training and test sets.}
\label{tab:experiments}
\end{table}

\begin{figure}[!h]
\centering
\includegraphics[width=1.0\textwidth]{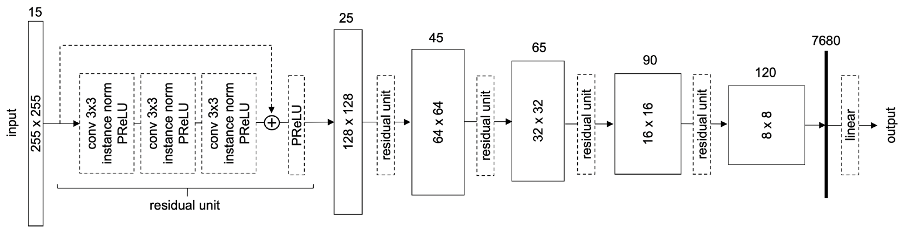}
\caption{Schematic diagram of the regression CNN used. The network was implemented in PyTorch using the ’Regressor’ module from the MONAI library \cite{Cardoso2022MONAI:Healthcare}. The architecture consists of 5 residual units which downsample the inputs by a factor of 2 via strided convolutions. Each residual unit consists of three convolutional layers (kernel size 3×3) followed by instance normalisation and a PReLU activation function. Skip connections are employed to bypass the convolutions. The network ends with a fully connected layer resizing the output from the residual blocks to a single value to which a linear activation function is applied.}
\label{fig:RegressionCNN}
\end{figure}

We trained the CNN shown in Fig \ref{fig:RegressionCNN} for the regression task using a mean square error loss function. We trained for 150 epochs, using Adam optimisation with a $10^{-4}$ learning rate, $10^{-4}$ weight decay, batch size of 2 and 0.1 dropout. 

\section{Results}
\subsubsection{MRI Simulator}
\begin{figure}[!h]
\centering
\includegraphics[width=1.0\textwidth]{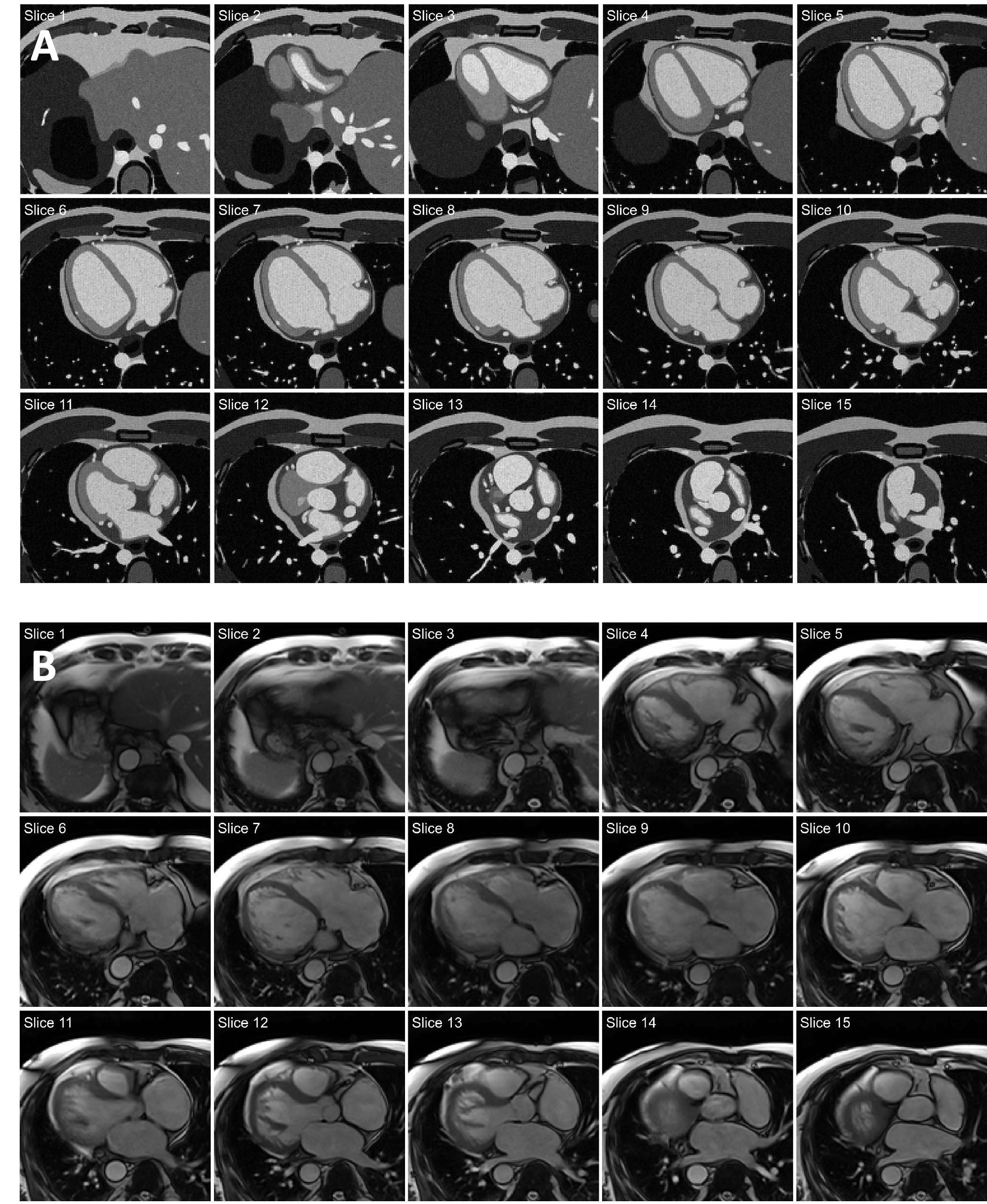}
\caption{Examples of stacks of axial cardiac images. A: Images simulated using the proposed framework. B: Patient images acquired in a 1.5T scanner.}
\label{fig:SimImas}
\end{figure}

The simulator was able to generate realistic sets of cardiac MR images according to the visual assessment of experts. All represented anatomical structures preserved their smooth, non-overlapping boundaries. A representative example is shown in Fig \ref{fig:SimImas}A. The simulation of each 3D cardiac image took approximately 2 h on a single CPU. By distributing the image generation process across 64 CPUs, we were able to simulate sets of 500 cardiac MR images in less than 16 h. Despite differences in contrast and anatomy, the simulated images compare well with the cropped patient images depicted in Fig \ref{fig:SimImas}B, showing similar cardiac structures in an equivalent anatomical context.

\subsection{Left Ventricular Myocardial Volume (LVMV) Estimation} With the introduced variations in cardiac anatomy, we were able to simulate images with varying realistic cardiac sizes and morphology.
VLVM in the simulated images was $102 \pm 19 cm^3$ (range: $55-163~cm^3$), compared to $123 \pm 60 cm^3$ (range: $43-242~cm^3$) from the segmentations of the patient data.

\begin{figure}[!h]
\centering
\includegraphics[width=1.0\textwidth]{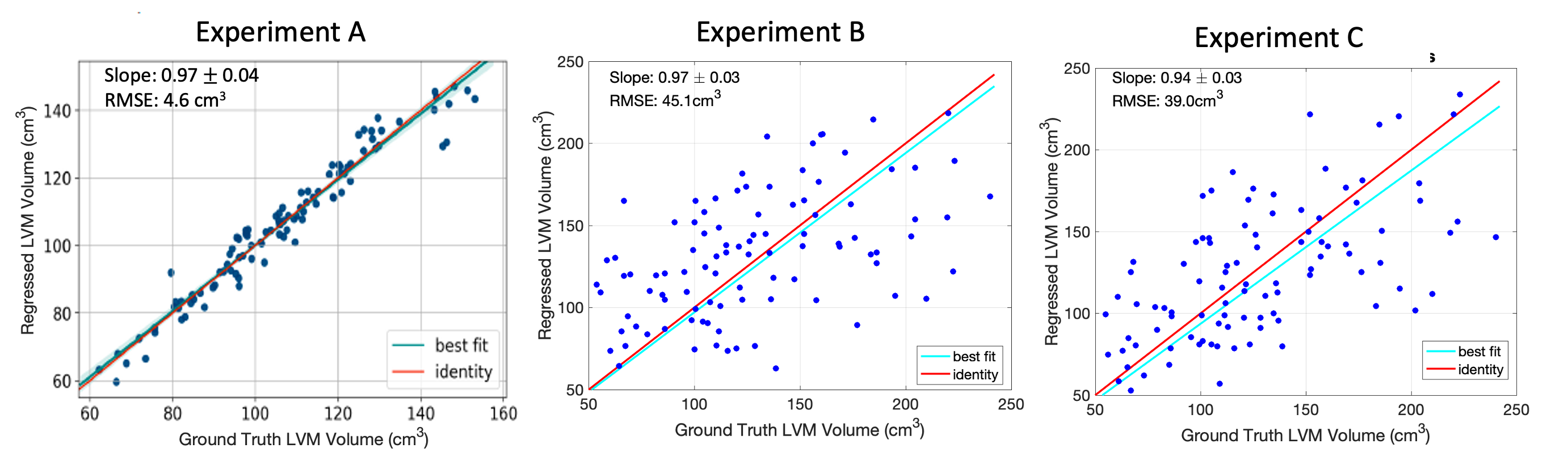}
\caption{Scatter plots on test data of LVMV estimation. From left to right, we depict the outcomes of experiments A, B and C respectively (see Table \ref{tab:experiments} for further details). In each case, the identity y=x line and the best fit line are also shown. We also indicate the values of the slope of the best fit line (going through the origin) and the root mean square error of the volume estimates in each experiment.}
\label{fig:LVVFits}
\end{figure}

The regression CNN was able to estimate LVMV, as shown in Fig \ref{fig:LVVFits}. LVMV estimates on simulated data were accurate and precise (Fig \ref{fig:LVVFits}a), when compared to LVMV estimates performed on patient images (Figs \ref{fig:LVVFits}b and c). Enhancing the training dataset with simulated data (Fig \ref{fig:LVVFits}c) led to a small decrease in the dispersion of the estimates (root mean square error (RMSE) of $39.0~cm^3$ instead of $45.1~cm^3$, whilst introducing a small drop in accuracy (best fit slope of $0.94 \pm 0.03$ vs $0.97 \pm 0.03$). The comparisons are performed against the data shown in Fig \ref{fig:LVVFits}b, where no simulated images were used for training.  

\section{Conclusions}
We present a detailed cardiac MRI simulator, capable of reproducing the main imaging features of cardiac MRI with a high anatomical detail. The proposed simulator is designed to allow the rapid and efficient generation of batches of cardiac MR images, with controlled variation in some parameters (anatomical and/or related to MR imaging). This makes the simulator ideal for the development and testing of machine learning applications. Suitable tasks include the training of supervised NNs which rely on large amounts of training data, such as NNs for image analysis tasks, as the LVMV estimation task presented here exemplifies. It is also well suited to train NNs for image reconstruction or image manipulation tasks, for which ground truth information is often not readily available. Finally, it offers a platform to study the robustness of existing NNs to controlled training distribution shifts or adversarial perturbations, and to address potential imbalances in existing datasets.

Others have also seen the potential of simulated MR images for machine learning. For example, Xanthis \textit{et al} \cite{Xanthis2021Simulator-generatedMRI} have generated synthetic MR images to train a neural network for a cardiac segmentation task. Their approach, however, is not specifically catered towards allowing the rapid parallel generation of MRI images whose parameters follow a specific statistical distribution. 

Despite its potential, the current implementation of the simulator is still a work in progress and suffers from some limitations that restrict the realism of the images it produces. For example, the simulator currently does not include: cardiac and breathing motion; the effects of water diffusion or blood flow; interactions between different proton pools; partial volume effects; inhomogeneities in $B_0$ or $B_1$ fields; variations in magnetic susceptibility or realistic transmit and receive coil sensitivities. These limitations explain the relative lack of success we achieved when enhancing the training of our LVMV regression network with simulated data (see Fig \ref{fig:LVVFits}). The LVMV regression network performed well on the synthetic data where tissue intensities are relatively flat and well defined, but poorly on real data, with its more complex intensity distribution. In this instance, adding simulated data to the training pool did not help the network's performance, but this is likely to change when more receive coil sensitivities are included in the simulation process.

Future work will test the usefulness of the proposed cardiac MRI simulator in other datasets and tasks. We will test the proposed LVMV estimation CNN on open access cardiac MRI databases with LV segmentations delineated by experts, such as the short-axis images available from the 2011 LV Segmentation Challenge \cite{LVSegChallenge2011}. The proposed simulator is also well suited for training networks for other tasks, such as the identification of anatomical structures in different MRI slices. It is currently being tested for this purpose.

The XCAT phantom setup we used allows for several variations in anatomy and properties, from variations in heart dimensions to alterations in LV thickness, which mimic both physiological variability and disease-induced remodelling. It currently does not allow changes in the anatomy of the other thoracic and abdominal organs present in the images or in the orientation of the heart relative to its surroundings, although patient data shows a large degree of variation in this latter parameter. The current version of XCAT phantom also does not permit the straightforward inclusion of focal pathology, such as myocardial scarring.

Despite its current shortcomings, we believe that the presented cardiac MRI simulator is a useful and attractive platform for the generation of large datasets of synthetic cardiac MR images in a controlled and simple manner. We expect that in the future these synthetic datasets will be used to train NNs and improve their robustness and explainability.




\section{Acknowledgments}
This work was supported by the British Heart Foundation Centre of Research Excellence at Imperial College London (RE/18/4/34215).
The authors would like to thank Abhishek Roy, Tommy Chen and Krithika Balaji for their contributions.

%
%
%
\bibliographystyle{splncs04}
\bibliography{references}

\begin{thebibliography}{10}
\providecommand{\url}[1]{\texttt{#1}}
\providecommand{\urlprefix}{URL }
\providecommand{\doi}[1]{https://doi.org/#1}

\bibitem{Benoit-Cattin2005TheSimulator}
Benoit-Cattin, H., Collewet, G., Belaroussi, B., Saint-Jalmes, H., Odet, C.:
  {The SIMRI project: A versatile and interactive MRI simulator}. Journal of
  Magnetic Resonance  \textbf{173}(1),  97--115 (3 2005).
  \doi{10.1016/j.jmr.2004.09.027}

\bibitem{Bittoun1984AMethod}
Bittoun, J., Taquin, J., Sauzade, M.: {A computer algorithm for the simulation
  of any Nuclear Magnetic Resonance (NMR) imaging method}. Magnetic Resonance
  Imaging  \textbf{2}(2),  113--120 (1 1984).
  \doi{10.1016/0730-725X(84)90065-1}

\bibitem{Cardoso2022MONAI:Healthcare}
Cardoso, M.J., Li, W., Brown, R., Ma, N., Kerfoot, E., Wang, Y., Murrey, B.,
  Myronenko, A., Zhao, C., Yang, D., Nath, V., He, Y., Xu, Z., Hatamizadeh, A.,
  Myronenko, A., Zhu, W., Liu, Y., Zheng, M., Tang, Y., Yang, I., Zephyr, M.,
  Hashemian, B., Alle, S., Darestani, M.Z., Budd, C., Modat, M., Vercauteren,
  T., Wang, G., Li, Y., Hu, Y., Fu, Y., Gorman, B., Johnson, H., Genereaux, B.,
  Erdal, B.S., Gupta, V., Diaz-Pinto, A., Dourson, A., Maier-Hein, L., Jaeger,
  P.F., Baumgartner, M., Kalpathy-Cramer, J., Flores, M., Kirby, J., Cooper,
  L.A.D., Roth, H.R., Xu, D., Bericat, D., Floca, R., Zhou, S.K., Shuaib, H.,
  Farahani, K., Maier-Hein, K.H., Aylward, S., Dogra, P., Ourselin, S., Feng,
  A.: {MONAI: An open-source framework for deep learning in healthcare}  (11
  2022). \doi{10.48550/arxiv.2211.02701},
  \url{https://arxiv.org/abs/2211.02701v1}

\bibitem{Howard2021a}
Howard, J., Zaman, S., Ragavan, A., Hall, K., Leonard, G., Sutanto, S.,
  Ramadoss, V., Razvi, Y., Linton, N.F., Bharath, A., Shun-Shin, M.J.,
  Rueckert, D., Francis, D., Cole, G.: {Automated analysis and detection of
  abnormalities in transaxial anatomical cardiovascular magnetic resonance
  images: a proof of concept study with potential to optimize image
  acquisition}. International Journal of Cardiovascular Imaging
  \textbf{37}(3),  1033--1042 (2021). \doi{10.1007/s10554-020-02050-w},
  \url{https://doi.org/10.1007/s10554-020-02050-w}

\bibitem{Kainz2019}
Kainz, W., Neufeld, E., Bolch, W.E., Graff, C.G., Kim, C.H., Kuster, N., Lloyd,
  B., Morrison, T., Segars, W.P., Yeom, Y.S., Zankl, M., Xu, X.G., Tsui, B.M.:
  {Advances in computational human phantoms and their applications in
  biomedical engineering - A topical review} (1 2019).
  \doi{10.1109/TRPMS.2018.2883437}

\bibitem{Kose2017BlochSolver:Sequences}
Kose, R., Kose, K.: {BlochSolver: A GPU-optimized fast 3D MRI simulator for
  experimentally compatible pulse sequences}. Journal of Magnetic Resonance
  \textbf{281},  51--65 (8 2017). \doi{10.1016/j.jmr.2017.05.007}

\bibitem{Litjens2019State-of-the-ArtAnalysis}
Litjens, G., Ciompi, F., Wolterink, J.M., de~Vos, B.D., Leiner, T., Teuwen, J.,
  I{\v{s}}gum, I.: {State-of-the-Art Deep Learning in Cardiovascular Image
  Analysis}. JACC: Cardiovascular Imaging  \textbf{12}(8P1),  1549--1565 (8
  2019). \doi{10.1016/J.JCMG.2019.06.009},
  \url{https://www.jacc.org/doi/10.1016/j.jcmg.2019.06.009}

\bibitem{Liu2017FastModel}
Liu, F., Velikina, J.V., Block, W.F., Kijowski, R., Samsonov, A.A.: {Fast
  Realistic MRI Simulations Based on Generalized Multi-Pool Exchange Tissue
  Model}. IEEE Transactions on Medical Imaging  \textbf{36}(2),  527--537 (2
  2017). \doi{10.1109/TMI.2016.2620961}

\bibitem{Segars2010}
Segars, W.P., Sturgeon, G., Mendonca, S., Grimes, J., Tsui, B.M.: {4D XCAT
  phantom for multimodality imaging research}. Medical Physics  \textbf{37}(9),
   4902--4915 (2010). \doi{10.1118/1.3480985}

\bibitem{Segars2018ApplicationBeyond}
Segars, W.P., Tsui, B.M., Cai, J., Yin, F.F., Fung, G.S., Samei, E.:
  {Application of the 4-D XCAT Phantoms in Biomedical Imaging and beyond}. IEEE
  Transactions on Medical Imaging  \textbf{37}(3),  680--692 (3 2018).
  \doi{10.1109/TMI.2017.2738448}

\bibitem{Stoecker2010High-performanceSimulations}
Stoecker, T., Vahedipour, K., Pflugfelder, D., Shah, N.J.: {High-performance
  computing MRI simulations}. Magnetic Resonance in Medicine  \textbf{64}(1),
  186--193 (7 2010). \doi{10.1002/mrm.22406},
  \url{https://onlinelibrary.wiley.com/doi/10.1002/mrm.22406}

\bibitem{LVSegChallenge2011}
Suinesiaputra, A., Cowan, B.R., Al-Agamy, A.O., Elattar, M.A., Ayache, N.,
  Fahmy, A.S., Khalifa, A.M., Medrano-Gracia, P., Jolly, M.P., Kadish, A.H.,
  Lee, D.C., Margeta, J., Warfield, S.K., Young, A.A.: {A collaborative
  resource to build consensus for automated left ventricular segmentation of
  cardiac MR images}. Medical Image Analysis  \textbf{18}(1),  50--62 (1 2014).
  \doi{10.1016/J.MEDIA.2013.09.001}

\bibitem{VaroquauxMachineFuture}
Varoquaux, G., Cheplygina, V.: {Machine learning for medical imaging:
  methodological failures and recommendations for the future} .
  \doi{10.1038/s41746-022-00592-y},
  \url{https://doi.org/10.1038/s41746-022-00592-y}

\bibitem{Xanthis2019CoreMRI:Cloud}
Xanthis, C.G., Aletras, A.H.: {coreMRI: A high-performance, publicly available
  MR simulation platform on the cloud}. PLOS ONE  \textbf{14}(5),  e0216594 (5
  2019). \doi{10.1371/journal.pone.0216594},
  \url{https://dx.plos.org/10.1371/journal.pone.0216594}

\bibitem{Xanthis2021Simulator-generatedMRI}
Xanthis, C.G., Filos, D., Haris, K., Aletras, A.H.: {Simulator-generated
  training datasets as an alternative to using patient data for machine
  learning: An example in myocardial segmentation with MRI}. Computer Methods
  and Programs in Biomedicine  \textbf{198},  105817 (1 2021).
  \doi{10.1016/j.cmpb.2020.105817}

\bibitem{Young2009}
Young, A.A., Frangi, A.F.: {Computational cardiac atlases: From patient to
  population and back}. Experimental Physiology  \textbf{94}(5),  578--596
  (2009). \doi{10.1113/expphysiol.2008.044081}

\end{thebibliography}

\end{document}